\documentclass[12pt,preprint]{aastex}
\newcommand{\vc}[1]{\mbox{\boldmath $#1$}}
\begin{document}
%\date{\today}
\title{Tests of AMiBA Data Integrity}

\author{
Hiroaki Nishioka\altaffilmark{1}, 
Fu-Cheng Wang\altaffilmark{2,3},
Jiun-Huei Proty Wu\altaffilmark{2,3}, 
Paul T.P. Ho\altaffilmark{1,4}, 
Chih-Wei Locutus Huang\altaffilmark{2,3}, 
Patrick M. Koch\altaffilmark{1}, 
Yu-Wei Liao\altaffilmark{2,3}, 
Kai-Yang Lin\altaffilmark{1,2}, 
Guo-Chin Liu\altaffilmark{1,5}, 
Sandor M. Molnar\altaffilmark{1}, 
Keiichi Umetsu\altaffilmark{1,3}, 
Mark Birkinshaw\altaffilmark{6}, 
Pablo Altamirano\altaffilmark{1}, 
Chia-Hao Chang\altaffilmark{1}, 
Shu-Hao Chang\altaffilmark{1}, 
Su-Wei Chang\altaffilmark{1}, 
Ming-Tang Chen\altaffilmark{1}, 
Chih-Chiang Han\altaffilmark{1}, 
Yau-De Huang\altaffilmark{1}, 
Yuh-Jing Hwang\altaffilmark{1}, 
Homin Jiang\altaffilmark{1}, 
Michael Kesteven\altaffilmark{7}, 
Derek Y. Kubo\altaffilmark{1}, 
Chao-Te Li\altaffilmark{1}, 
Pierre Martin-Cocher\altaffilmark{1}, 
Peter Oshiro\altaffilmark{1}, 
Philippe Raffin\altaffilmark{1}, 
Tashun Wei\altaffilmark{1}, 
Warwick Wilson\altaffilmark{7}}
\email{nishioka@asiaa.sinica.edu.tw}
\altaffiltext{1}{Academia Sinica Institute of Astronomy and
Astrophysics, P.O. Box 23-141, Taipei 10617, Taiwan}
\altaffiltext{2}{Department of Physics, National Taiwan University, 
Taipei 10617, Taiwan}
\altaffiltext{3}{LeCosPA, National Taiwan University, 
Taipei 10617, Taiwan}
\altaffiltext{4}{Harvard-Smithsonian Center for Astrophysics, 60 Garden
Street, Cambridge, MA 02138, USA}
\altaffiltext{5}{Department of Physics, Tamkang University, 251-37
  Tamsui, Taipei County, Taiwan}
\altaffiltext{6}{University of Bristol, Tyndall Avenue, Bristol BS8 1TL, UK}
\altaffiltext{7}{Australia Telescope National Facility, P.O.Box 76, 
Epping NSW 1710, Australia}
\begin{abstract}
We describe methods used to validate data from the Y.T. Lee Array 
for Microwave Background Anisotropy (AMiBA), an interferometric 
array designed to 
measure the Sunyaev-Zel'dovich effect and the anisotropy of the 
Cosmic Microwave Background (CMB). We perform several statistical 
tests on data from pointed galaxy cluster observations taken in 2007
and noise data from long-term blank sky observations and 
measurements with the feeds covered by the absorbers.
We apply power spectrum analysis, cross power 
spectrum analysis among different outputs with different time lags 
in our analog correlator, and sample variance law tests to noise data.
We find that (1) there is no time variation 
of electronic offsets on the time scale of our two-patch 
observations ($\sim 10$ minutes); (2) noise is correlated by less 
than 10\% between different lags; and (3) the variance of noise 
scales with the inverse of time. To test the Gaussianity of the 
data, we apply Kolmogorov-Smirnov (K-S) tests to cluster data, and 
find that a 5\% significance level efficiently detects data sets 
with known hardware problems without rejecting an excess of 
acceptable data.
We also calculate third- and fourth-order moments 
and cumulants for the noise residual visibilities and find that 
about 95\% of our data are within the 99\% confidence regions of 
Gaussianity. 
\end{abstract}
\keywords{cosmic microwave background---cosmology:observations---
methods:data analysis}
\section{Introduction}
Measurements of the Cosmic Microwave Background (CMB) on arcminute 
scales reveal the large-scale structure of the Universe and 
probe the gas properties of galaxy clusters through the 
Sunyaev-Zel'dovich effect (SZE). Since the CMB signals on such small 
angular scales are quite weak, long integrations are necessary to 
achieve detections. Consequently the noise properties and system 
stability become crucial problems for CMB projects. Instability and 
non-Gaussianity of the noise often arise from hardware issues or 
unexpected noise sources in the system, which may cause systematic 
errors in the measurements and lead to incorrect scientific 
interpretation. Many CMB projects therefore expend significant 
effort on testing the noise properties of their data 
\citep[e.g.,][]{Beretal2000,Padinetal2001,Haletal2002,Graingeetal2003,Hinshawetal2003,Smithetal2004,Kuoetal2004,
Savageetal2004,Rubinoetal2006,Watsonetal2003,Wuetal2001,Jarosiketal2007,Wuetal2007,Muchoetal2007}. 

This paper is one of a series of papers reporting the first 
results from the Y.T. Lee Array for Microwave Background Anisotropy 
(AMiBA) \citep{AMiBA_Ho}. AMiBA is an interferometric array 
designed to measure the anisotropy of the Cosmic Microwave Background (CMB) 
and the SZE which has the dual-channel
86-102 GHz operating frequency \citep{AMiBA_Chen}. 
The site is located on Mauna Loa, 
Hawaii at 3396m altitude. Seven 60-cm antennas are
mounted on the 6-m configurable platform in the compact 
configuration in the current setup which is capable of an expansion
to 19 elements. Operation with 13 elements is planned 
to start in early 2009. The first run of the science operations 
was carried out in 2007 and a total of six galaxy clusters are 
detected through the SZE. The analysis results of these clusters
are reported in a series of papers 
\citep[][AMiBA cluster papers hereafter]{AMiBA_Wu,AMiBA_Umetsu,AMiBA_Koch_hub,AMiBA_Huang}.

The aim of this paper is to test the integrity of our data 
by performing statistical tests of the noise in our system, 
and thus to provide the basis on which the results of the 
AMiBA cluster papers are founded. Some other possible sources of systematic 
error are discussed in other papers. Detailed 
descriptions of the system performance and calibration are discussed 
in \citet{AMiBA_Lin}. Pointing accuracy and antenna misalignment 
are discussed in \citet{AMiBA_Koch_mou}. The level of contamination of 
SZE measurements from primary CMB fluctuations and radio sources or 
other foregrounds is discussed in \citet{AMiBA_Liu}. The overall   
data flagging is described in \citet{AMiBA_Wu}. 

This paper is organized as follows.
In section \ref{noise}, we discuss the statistical properties of the 
noise data. In section \ref{gau},
the Gaussianity tests of our data from cluster observations 
are described. Section \ref{con} is devoted to the Conclusions.

\section{Statistical tests of the noise in our system}\label{noise}

In this section we describe the noise properties of our system from 
the long-term noise data. The data were taken in two different ways. 
First, we took data with the feeds covered by absorbers that block 
all signals, for periods of one day to five days. 
Second, we made blank sky observations lasting about one hour to eight
hours at night (after sunset to before sunrise), 
where the telescope was parked pointing to the zenith 
(declination +19.5 degrees) --- we confirmed that no strong 
source passes through the field of view. 
Multiple such datasets have been taken and 
similar characteristics were found in all cases. Representative 
results are shown below. Furthermore, we find no significant 
differences in the results of the statistical analyses described 
below between data taken in these two different ways. 

AMiBA is an interferometer with 4-lag analog correlators which 
output real-number correlation signals. We denote the time-ordered 
outputs from our correlators at time $t$ as 
$\vc{c}(t)=\{c_i(t):i=1,4\}$ in units of counts from the A/D 
converter with the time interval of 0.452 seconds
where $i=1,...,4$ is the lag number \citep[see][]{Lietal2004}.
We define the power spectrum as 
 \begin{equation} 
   P_{ij}(\nu)={\hat{c}_i(\nu)\hat{c}_j^*(\nu)\over T}, 
   \label{powereq} 
 \end{equation} 
 where $\nu$ is the frequency, $\hat{c}$ is the Fourier transform of 
 $c(t)$, and ${}^*$ denotes complex conjugation. The power spectrum 
 is normalized by $T$, the total length of data. 

\subsection{Long time scale variation of the electronic offset}

\begin{figure}
\epsscale{.80}
\plotone{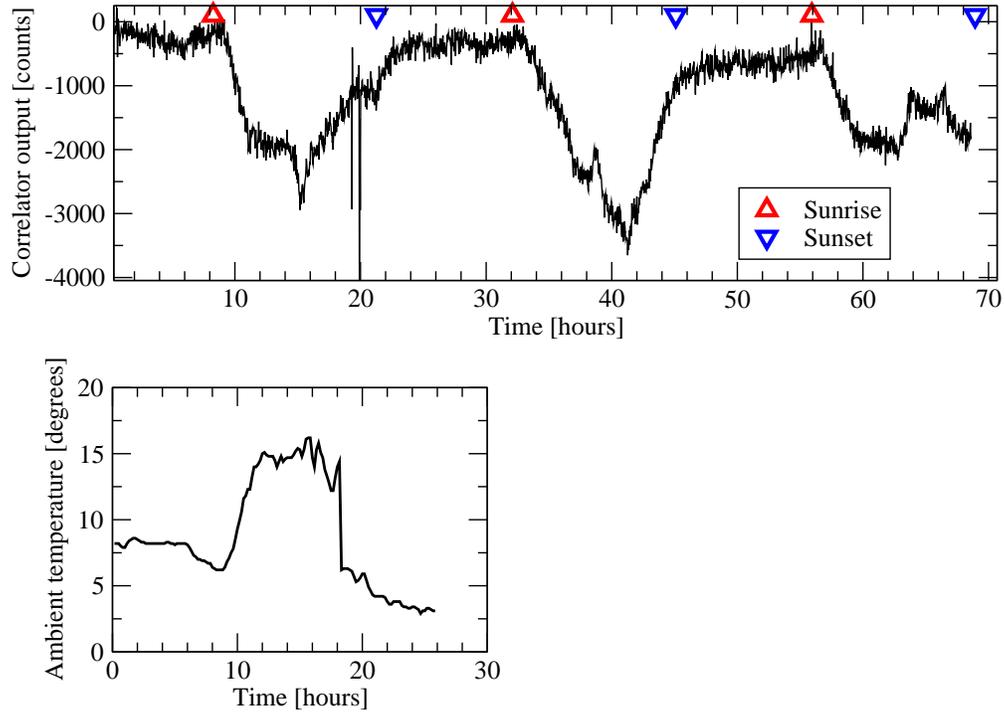}
\caption{The variation of the correlator outputs over about three 
days of data. Average counts over two-minute periods are shown as 
a function of time. A strong diurnal pattern is seen, with rapid 
changes at the times of sunrise and sunset (marked by the 
red triange-up and blue triangle-down, respectively).
The lower panel shows ambient temperature variations for about 
one day monitored over the same period.}
\label{timeval}
\end{figure}

Figure \ref{timeval} shows the time variation of the correlator 
outputs as a function of time over three days of absorber data.
The correlator outputs are averaged over two minutes to 
show the long-period variations more clearly. Although there is no 
astronomical signal in this data, the correlator outputs show 
time-variable and non-zero electronic offsets. Our data analysis 
procedure subtracts these offsets by adopting a two-patch procedure, 
where the telescope first tracks a science target for some time, and 
then moves to a blank-sky patch that trails the science target and 
follows the same set of mount positions \citep{AMiBA_Wu}. 
This ensures that electronic 
offsets and ground pick up signals that depend on mount position are 
accurately subtracted, provided that the time-scale of variation of 
these contaminating signals is sufficiently slow \citep{AMiBA_Wu}. 

Figure \ref{timeval} shows a strong diurnal variation in the 
correlator outputs which parallels ambient temperature variations. 
The most rapid changes in correlator signal occur at the sunrise and 
sunset, when the temperature changes quickly. At night, when the 
temperature variation is slow, the outputs are relatively stable. 
We avoid making observations near sunrise and sunset to exclude 
systematic errors from such rapid system changes. 
The results from the observations in 2007 reported 
in the AMiBA cluter papers are based on data taken only at night. 
We also carried out long-term tracking of the bright planets 
to examine the overall stability of the system including the effects 
from the atmosphere. From this test we concluded that 
gain and phase calibrations every two to three hours 
are necessary \citep{AMiBA_Lin}.

It is important to quantify the time scales of correlator variations 
to define the switching time for our two-patch method. We discuss 
this issue in the next section, through Fourier analysis. 

\subsection{Noise Power Spectrum}
If the time scale of the variation of the offsets is shorter than 
the switching interval in our two-patch procedure, then our data 
will be strongly contaminated by an error signal. To examine the 
time variation of the offsets, we perform a power spectrum analysis 
(the case of $i=j$ in equation (\ref{powereq})) of the noise data. 

\begin{figure}
\epsscale{.80}
\plotone{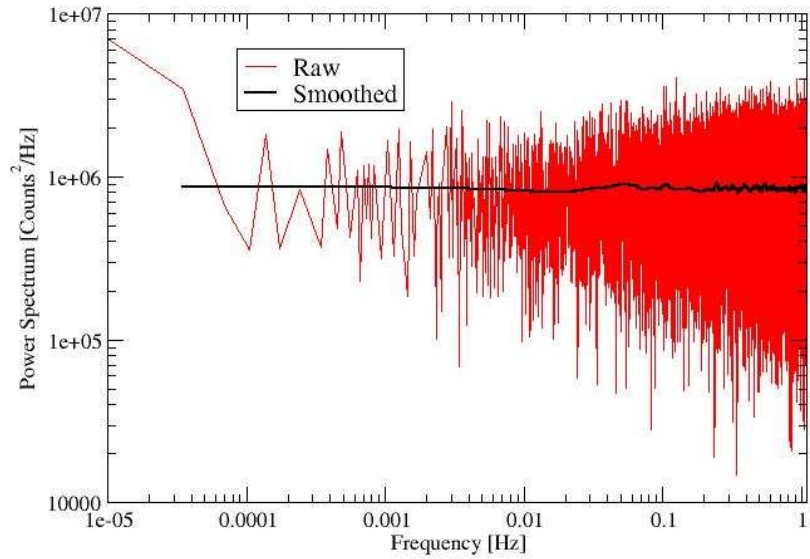}
\caption{The noise power spectrum from an eight-hour absorber 
data. The red and black curves represent the raw and 
smoothed power spectra, respectively. The spectra are indicative of 
white noise over the frequency range $0.0001$~to $1$~Hz. To reduce 
the sample variance, three sets of 8-hour spectra are averaged.} 
\label{power}
\end{figure}

Figure~\ref{power} shows the measured noise power spectrum 
for eight hours of data taken at night. The red curve 
shows the raw power spectrum, while the black curve shows 
the power spectrum smoothed by a Gaussian kernel with 
$\sigma=0.005$~Hz to show the characteristics of the data more 
clearly. To reduce the sample variance, three sets of 8-hour
spectra are averaged. The spectrum has white-noise form over 
frequencies between 
$10^{-4}$~and 1~Hz, with an increase in power at lower frequency 
because of slow variations of the offsets. Based on these results, 
we expect that two-patch data taken with an interval of less than 
600~sec (typical for our observing procedure) will not be 
contaminated by variations of the electronic offsets. 
\subsection{Noise Correlation among Lags}\label{noisecorr}
Each correlator output dataset consists of a time-ordered series of 
four numbers corresponding to the correlated signals from each of 
the four time lags. These four output numbers are transformed to the 
real and imaginary parts of two complex visibilities, representing 
the upper and lower frequency channels, in the first stages of 
data processing \citep{AMiBA_Wu}. Since this process is a linear 
transformation, noise correlation between different lag outputs 
leads to non-zero non-diagonal components of the noise correlation 
matrix, and this feeds through to the covariance matrix of the 
visibilities and error estimates. To examine how the noise from 
different lags correlates, we compute the cross power spectrum 
between lags. 

\begin{figure}
\epsscale{.80}
\plotone{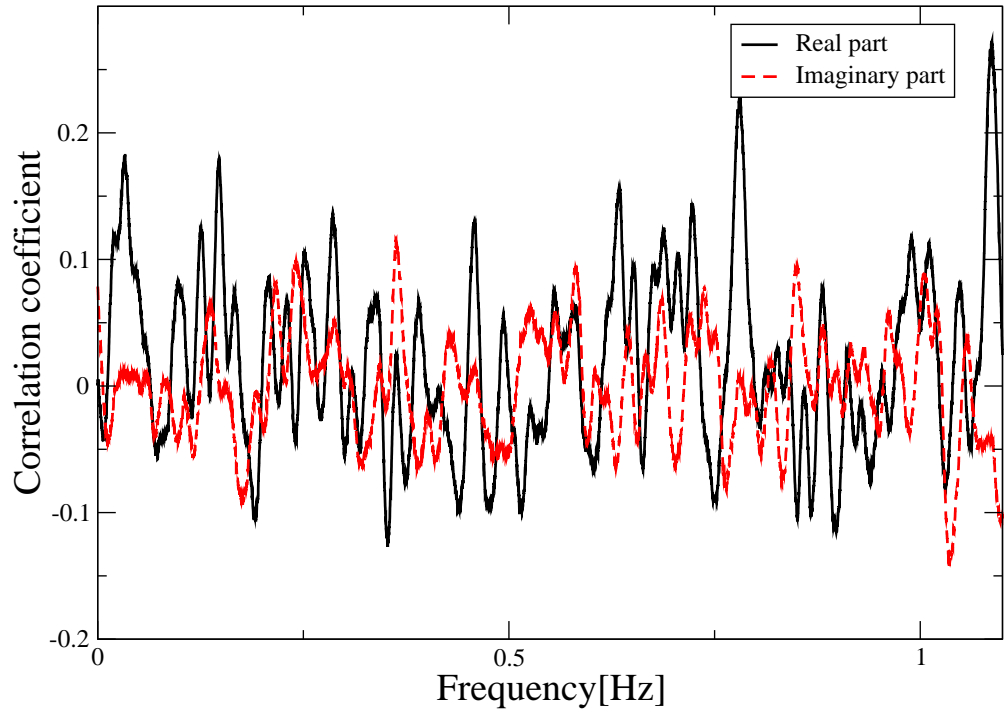}
\caption{The correlation coefficient among lags(the adjacent lag,
$i=1,j=2$) fron the blank sky data. 
The black-solid and red-dashed curves 
represent real and imaginary parts, respectively. 
The real part fluctuates around a positive value $\sim 0.03$.
There is no significant frequency dependence of $r$ and 
the level of correlation between lags is generally less than 10\%.}
\label{cpower}
\end{figure}

The strength of the correlation is quantified by the correlation 
coefficient defined by 
\begin{equation} 
r={P_{ij}(\nu)\over \sqrt{P_{ii}(\nu)P_{jj}(\nu)}}, 
\end{equation} 
where $r$ is a complex number. The real part of $r \rightarrow 1$ 
if lags $i$ and $j$ show perfect correlation. If there is no 
correlation, then $r \rightarrow 0$. Figure~\ref{cpower} shows the 
real and imaginary parts of the correlation coefficient between 
the adjacent lags, $i=1,j=2$. We do not find any different 
characteristics of the cross correlation between distant lags
(e.g. $i=1,j=3$ and $i=1,j=4$). 
To see the cross-correlation properties clearly, we 
apply smoothing by a Gaussian kernel with $\sigma=0.05$~Hz. 

We see that the imaginary part of $r$ fluctuates around zero while 
the real part fluctuates around a positive value $\sim 0.03$, which 
suggests the existence of correlated noise between lags. However, 
we find no significant frequency dependence of $r$, and it is clear 
from Fig.~\ref{cpower} that the level of correlation between lags is 
generally less than 10\%. We note that there are some exceptions to 
this limit, where stronger correlation is found (sometimes more than 
20\%), but this is a characteristic of correlators with temporary 
hardware problems where the data would not be used in later analysis. 
\subsection{Sample Variance Law test}
In our science observations we rely on the reduction of the noise 
level through long integrations. The rate at which the noise reduces 
in time is an important indicator of the effectiveness of the AMiBA 
system, and we would hope that noise fluctuations decrease in time 
following the sample variance law. We test this by calculating 
whether the variance of the noise power spectrum decreases as the 
length of the data chunk used for the spectrum increases. We apply 
the test by\\ 
(i) dividing the raw datasets of the noise 
    into short chunks with different lengths\\ 
(ii) calculating the power spectrum for each chunk\\ 
(iii) calculating the variance of the chunk power spectra over 
      the entire dataset\\ 
(iv) comparing the variances from different chunk lengths.\\ 
The variance is defined by 
\begin{equation} 
  {\rm variance} = {1 \over N} \sum_i^N P_{\rm chunk}^2(i,\nu) 
    -\left[ {1 \over N} \sum_i^N P_{\rm chunk}(i,\nu)\right]^2 \quad , 
\end{equation} 
where $P_{\rm chunk}(i,\nu)$ is the power spectrum of the $i$-th 
chunk and $N$ is the number of chunks. 

\begin{figure}
\epsscale{.80}
\plotone{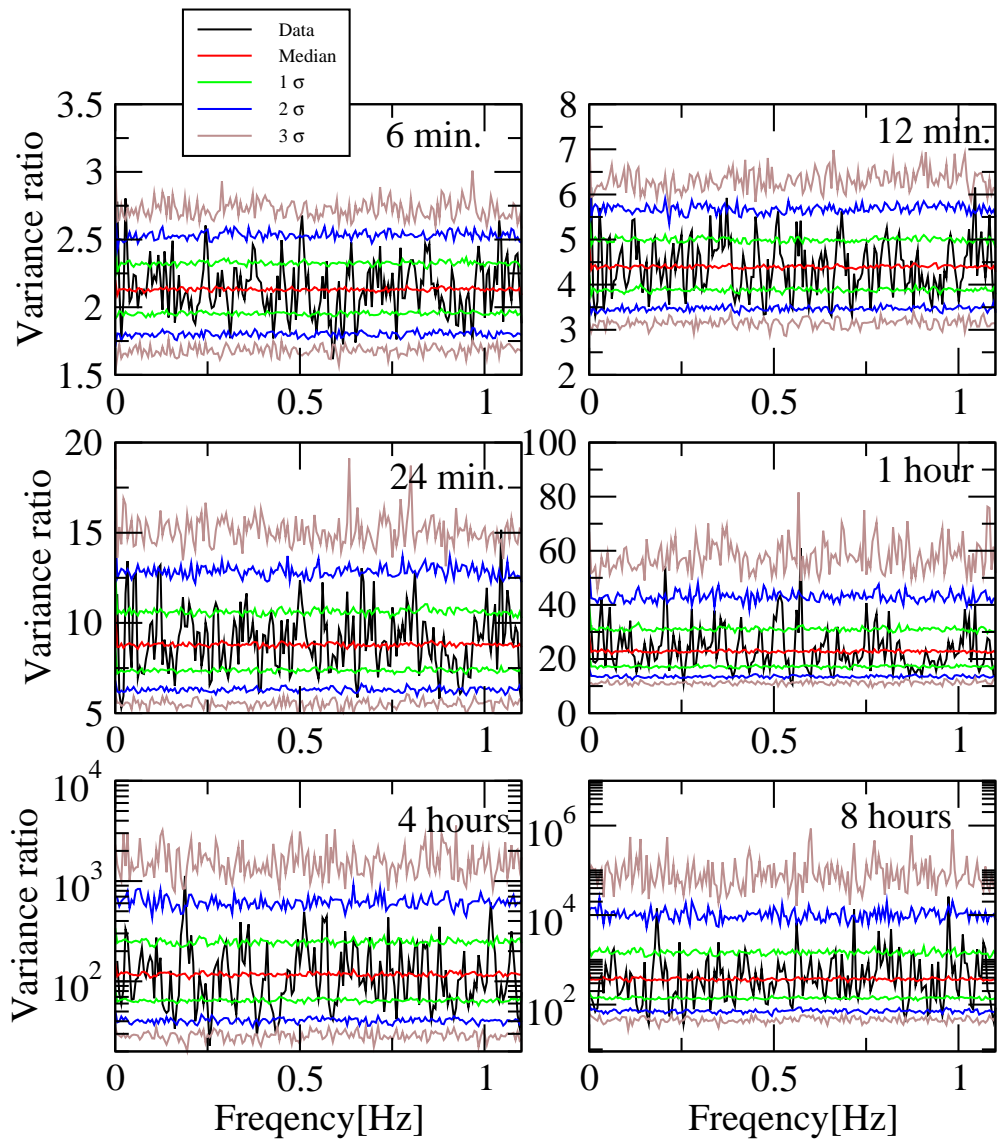}
\caption{Variance ratios of 3-minute chunk power spectrum 
to the 6-minute to 8-hour chunk spectra from the absorber data. 
The black curves represent the variance ratios, and 
the green, blue and brown lines represent the 1, 2 and~3$\sigma$ 
confidence levels derived from 500 Monte-Carlo simulations assuming 
Gaussian random noise with the same noise power as our data. The red
lines represent the medians from the simulations.} 
\label{ratios}
\end{figure}

Figure~\ref{ratios} shows the variance ratio, which is the variance 
of the power spectra for 3-minute chunks divided by the variances 
for 6 minutes to 8 hours-chunks. In this analysis
24 hours data (made up from three sets of eight-hour 
datasets taken at night) is used.
The black curves represent 
the measured variance ratio spectra. It is seen that the variance ratios 
fluctuate around the median from 500 Monte Carlo simulations 
(red lines) as expected if the chunk power follows the sample variance law. 
The green, blue and brown lines 
represent the 1, 2 and~3$\sigma$ confidence levels derived from the
simulations. In the simulations we assume Gaussian random noise with a flat 
power spectrum with the same level of power as the data. 
The results are consistent with the Monte Carlo simulations within 
3-$\sigma$ confidence limits, except for a few data points that 
we can trace to spurious correlator outputs that have negligible effect on our 
analyses or science results. 
The integration times of the observed galaxy clusters are 
five to eleven hours \citep{AMiBA_Wu}. The reduction of noise 
with integration demonstrated here assures us of the 
reliability of our cluster results.

For our observations of galaxy clusters, the contribution 
from the brightest contaminants in the night sky, the Moon and the planets,
is crucial. We also investigate it here using the measured primary 
beam pattern and the simulated correlator data \citep{AMiBA_Wu}.
In the 2-patch observing strategy, when the offset direction of the 
Moon is perpendicular to the baseline by 10 and 40 degrees
and the two-patch link direction is also perpendicular to the baseline,
the attenuation in Moon signal is by an order of $10^{-5}$ and $10^{-6}$ 
respectively. Such attenuation is mainly from the attenuation 
of the primary beam.
When the offset direction is along the baseline direction
and the two-patch link direction is also aligned with the baseline,
attenuation of at least one extra order appears due to the
frequency-band smearing effect of the correlators.

qWe have carefully checked the angular separation between 
these contaminants and our cluster targets,
and found that it was at least 10 degrees, mostly beyond 50 degrees.
We finally verified that after the flagging and the integration 
over the hours that may be from various days,
the contribution from either the Moon or the planets is at 
least two orders below our cluster signals.

\section{Gaussianity of the data}\label{gau}
If the data are non-Gaussian in their statistics, then the data are 
likely to contain systematic errors, and the errors are likely to be 
badly estimated. In this section we apply several statistical tests 
to cluster SZE data to verify the Gaussianity of our system noise in 
the lag and visibility data.  
\subsection{Gaussianity tests in correlator outputs}
Since our science observations are based on the two-patch method 
described above, the data from each patch should be a constant 
signal plus noise. If the noise follows a Gaussian distribution, 
all our data also should follow a Gaussian distribution centered 
on the astronomical signal. To test the Gaussianity of our data, we 
adopt a Kolmogorov-Smirnov (K-S) test which is commonly used to 
judge whether two distributions are different at a certain 
significance level. In the K-S test, we measure the maximum value 
of the absolute difference between two cumulative distribution
functions,
\begin{equation}
D=\max_{-\infty<x<\infty}  \left|S_{\rm data}(x)-S(x)\right|,
\end{equation}
where $S_{\rm data}(x)$ is the cumulative distribution of the 
correlator outputs. The cumulative distribution $S(x)$
which is to be compared with the data is derived from the 
Gaussian distribution centered at the 
mean signal and with a standard deviation estimated from the rms 
deduced from data during observations of a single patch.
The maximum distance $D$ is a measure of how much different 
the two distributions are. The low significance level 
is induced by the large value of $D$ and means that the 
difference between two distributions is significant.

\begin{figure}
\epsscale{.80}
\plotone{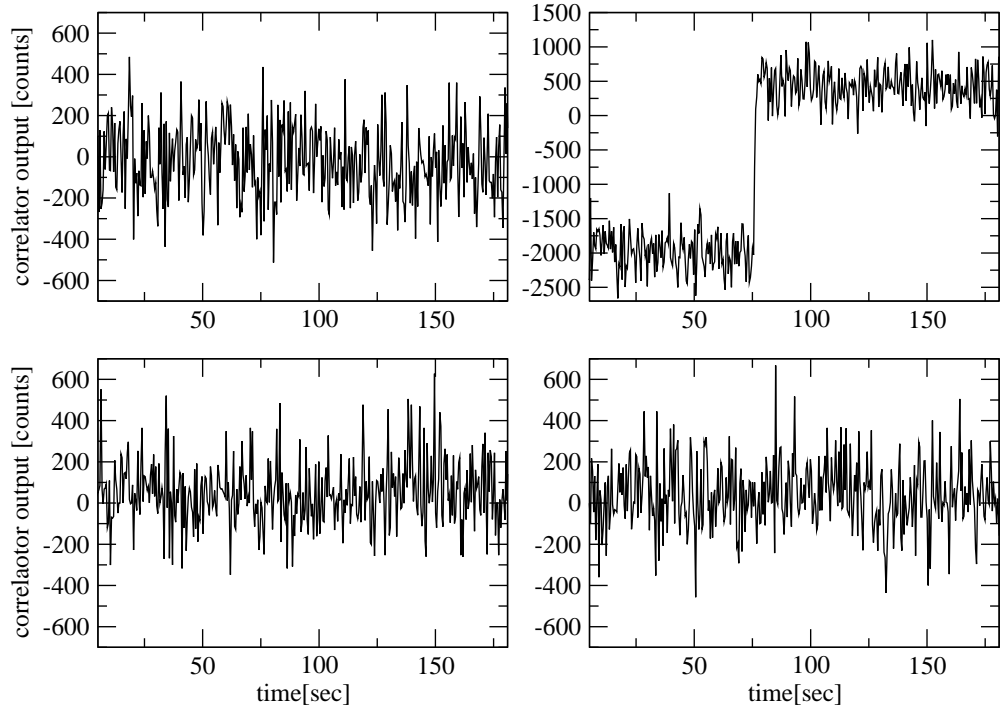}
\caption{Examples of correlator outputs as a function of time 
 for one patch of a two-patch cluster observation. The top left panel 
 an output that passes the K-S test. The other panels are examples of 
 outputs that fail the K-S test for obvious (top right) or more 
 subtle (bottom left and bottom right) reasons.} 
\label{ks}
\end{figure}

We apply this test to the data and find that our K-S tests 
at a 5\% significance level can successfully detect correlators 
that are known to provide bad 
data because of hardware problems. The tests with lower significance
levels reduce a success rate in detection of the problems.
Apart from these known bad 
correlators, more than 90\% of the data pass the K-S tests at the 
5\% significance level. Figure~\ref{ks} shows example correlator 
outputs for one patch of observation. The upper left panel shows an 
output lag dataset that passes the K-S test. The other panels are 
examples where the outputs fail the K-S test at the 5\% level. 
While the example shown in the top right panel is obvious, the 
examples in the lower two panels are less clear, although all three 
datasets were known to have hardware problems at the time that these 
data were taken. (The problem was temporary and quickly fixed). 
This suggests that K-S tests at a 5\% level provide a useful 
diagnostic for hardware problems. In our analysis procedure we 
remove all datasets identified by such a K-S test. 
\subsection{Gaussianity tests in visibilities}
We also apply Gaussianity tests to the visibilities. Since the 
visibilities from cluster observations include signal and noise, 
we must subtract the astronomical signals to investigate the residual 
noises. In our analysis we subtract the mean of the visibilities, 
and investigate the third- and fourth-order moments and cumulants 
from the residuals. We analyze the visibility data for the real and imaginary 
parts of the lower and upper frequency bands separately. We find 
that about 95\% of our data from the six clusters described by 
\citet{AMiBA_Wu} are consistent with Gaussianity within 99\% confidence 
limits as derived from 500 Monte-Carlo simulations (more realizations
do not change the results). Although this result implies that there 
is a small level of non-Gaussian noise present, 
this has no important effect on the statistics.
\section{Conclusions}\label{con}

We have performed several statistical tests to noise data from blank 
sky observations and the measurements with the feeds covered by the 
absorbers, and observations of galaxy clusters to demonstrate the 
integrity of AMiBA data. The noise data has been analyzed through 
their power spectra, lag-lag cross-correlations, and sample variance law. 

From the power spectrum analysis we have found that there are no 
significant time variations of the electronic offsets on time scales 
shorter that the time scale of our two-patch observing 
procedure. This demonstrates that our science data is free of 
systematic errors introduced by fast system drifts. 

Cross power spectrum analysis has been applied to examine the noise 
correlation between different lags. This revealed a weak 
correlation, at the level of less than 10\%. 

To examine whether the noise decreases over long integrations, we 
have applied the sample variance law test to long-duration blank-sky 
data. The data were divided into several chunks with different 
lengths in time and the variances of the power spectra were 
calculated and compared. We showed that the fluctuations of chunk 
power follow the sample variance law to the time scales of the
integration times of galaxy clusters to within the errors estimated 
from Monte Carlo simulations assuming that the data are scattered 
only by Gaussian random noise with the same power. 

To test the Gaussianity of our data we applied statistical 
tests to the lag and visibility data. We found that K-S tests with 
5\% significance can clearly detect lag and visibility data known to 
be bad because of hardware issues. This led us to adopt a K-S test 
at this level to check all AMiBA data for hardware problems. 
Except for such known bad lag data, more than 90\% of 
the data passes such K-S tests. 

We have also performed Gaussianity tests on the visibility data. 
We calculated third- and fourth-order moments and cumulants of the 
residual noise visibilities, and found that about 95 \% of our data 
are within the 99\% confidence region of Gaussianity derived from 
the Monte Carlo simulations. The apparent residual level of 
non-Gaussianity (4\%) is not expected to affect the analysis 
of AMiBA data. 

\acknowledgments
We thank all the members of the AMiBA team for their hard work.
We thank Katy Lancaster and Tzihong Chiueh for useful
comments and discussions. We thank anonymous referee for useful
suggestions to improve the original manuscript.
Support from the STFC for MB is also acknowledged.

\clearpage

\end{document}